\documentclass[aps,amsart
amsproc,article,elsarticle,emulateapj]{revtex4}
\usepackage{bm}
\usepackage{graphicx,amsmath,latexsym,amssymb}
\usepackage{xcolor}
\usepackage{sistyle}
\usepackage{pstricks,pst-grad}

\usepackage{pstricks,pst-grad}

\begin{document}
	
	\title{Collapsing dynamics of attractive Bose-Einstein condensates in random potentials}
	
	\author{Madjda Kamel and Abdel\^{a}ali Boudjem\^{a}a}
	\affiliation{Department of Physics,  Faculty of Exact Sciences and Informatics, Hassiba Benbouali University of Chlef P.O. Box 78, 02000, Ouled Fares, Chlef, Algeria.}
	\email {a.boudjemaa@univ-chlef.dz}
	
	
	\begin{abstract}
          We study the stationary and dynamical properties of three-dimensional trapped Bose-Einstein condensates with attractive interactions 
               subjected to a random potential. To this end, a variational method is applied to solve the underlying Gross-Pitaevskii equation. 
               We derive analytical predictions for the energy, the equilibrium width, and evolution laws of the condensate parameter.  
               The breathing mode oscillations frequency of the condensate has been also calculated in terms of the gas and disorder parameters.
               We analyze in addition the dynamics of collapse from the Gaussian approximation. 
               Surprisingly, we find that the intriguing interplay of the attractive interaction and disorder effects leads to prevent collapse of the condensate.

	\end{abstract}

	\maketitle
	
	\section{Introduction} \label{Intro}
	
The successful realization of Bose-Einstein condensates (BECs) of ultracold atoms in 1995 \cite{Rb,Na} has led to the appearance of advanced experimental and theoretical techniques, 
enabling the investigation of unexplored quantum phenomena and unique states of matter. 

        In the presence of repulsive interactions, a trapped BEC is always stable while attractive interactions give rise to unstable condensate as long as the number of particles is above 
	a critical value below which the condensate is in a metastable state. 
	The mean-field attraction brings the BEC into a region where the quantum fluctuations become important as the negative interaction  energy becomes 
	comparable with kinetic energy. Moreover, in the presence of attractive interaction, the ground-state contracts and the density increases rapidly \cite{Hulet1,Dodd}.
	Near the stability limit, self-attraction can overwhelm the repulsion, causing the condensate to collapse. 
	A direct observation of such a collapse of a BEC with attractive interactions has been reported in \cite{Hulet1,Hulet}.
	It has been found also that the condensate undergoes many cycles of growth and collapse before a stationary state is reached \cite{Stoof}.
	Note that homogeneous Bose gas with attractive interactions has been shown to be unstable. 
	Furthermore, the collapse of BECs with spin-orbit coupling  driven by both cubic and quintic  self-attraction has been investigated in \cite{Sherman1, Sherman2}.

	On the other hand, disorder exists to some extent in all natural media, and even a small amount of disorder may produce large and counterintuitive effects. 
       There are several methods to create disorder in BEC. 
	Common techniques are: speckle lasers, magnetic field gradients, random potentials and incommensurate laser beams \cite{Asp,Modu}. 
	Ultracold atomic gases offer appealing possibilities to study the physics of disordered quantum systems, as many parameters, 
        including disorder and interactions, may be controlled in these systems. 
	Experimentally, the dirty boson problem was first studied with superfluid  helium in aerosol glasses (Vycor) \cite{MPA,BCC,JDR}, 
	and theoretically it was studied  for the first time by Huang and Meng \cite{HM} in 1992 within the Bogoliubov theory \cite{bog} for weak disorder.
	
	In the past few years, dipolar and nondipolar Bose-condensed gases with repulsive interactions in disorder potentials have been extensively explored both theoretically (see e.g.  \cite{HM,Gior,Flac,jk,Boudj0,Boudj1,Boudj2,Boudj3}) and experimentally (see e.g. \cite{Asp,Modu,Pasi,Deiss,Nagler}). 
	Yet, less is known on BECs with attractive interactions and subject to a well-controlled disorder. 
        Experimental and numerical studies of one-dimensional (1D) BECs with attractive interactions in a random potential have shown 
	evidence of a bright soliton with an overall unchanged shape, but a disorder-dependent width \cite{Akk,Yong}.
	Moreover, subdiffusive spreading and localization landscape of a BEC with attractive interactions in 1D speckle potentials have been examined in \cite{Bao,Fill}.
	However, trapped BECs with attractive interactions in 3D random potentials remain largely unexplored.
	Studying attractive disordered BECs may  provide valuable insights to understand the behavior of nonlinear systems in the presence of randomness.

	The aim of this paper is then to investigate the equilibrium and dynamical properties of harmonically confined 3D  BECs with attractive interactions subjected
        to a 3D random potential made by a laser speckle. We look in particular at how the intriguing competition between attractive interactions and disorder alter the stability of the system 
        and the dynamics of collapse. 
	Within the realm of a variational method based on a Gaussian ansatz, we solve the Gross-Pitaevskii equation (GPE) which is a nonlinear Schr\"odinger equation with a disorder potential, 
        and calculate the critical number of particles above which the system is stable.
	The width and the breathing modes frequency of the condensate are also determined in terms of the gas and disorder parameters.
	 We analyze in addition  the dynamics of collapse from the Gaussian approximation.

	\section{Model}
	
	We consider a harmonically trapped BEC with attractive interactions subjected to a 3D disorder potential. The dynamics of the system is governed by the GPE:
	\begin{equation} \label{GPEd}
		i\hbar\frac{\partial\phi({\bf r},t)}{\partial t}=\bigg(\frac{-\hbar^2}{2m}\Delta+V_{\text{ext}}+U_{\text{dis}}-g|\phi|^2\bigg)\phi({\bf r},t),
	\end{equation}
	where $V_{\text{ext}} ({\bf r})=m \omega_0^2 r^2/2$ is the external potential, $m$ is the atomic mass, $\omega_0$ is the trap frequency,
	$g=4\pi\hbar^2 a/m$ is the coupling constant with $a$  being the $s$-wave scattering length. 
	The disorder potential $U_{\text{dis}} ({\bf r})$ is described by vanishing ensemble averages $\langle U_{\text{dis}}({\bf r}) \rangle =0$,
	and $\langle U_{\text{dis}} ({\bf r}) U_{\text{dis}}({\bf r'}) \rangle=R({\bf r}- {\bf r'})$, where $ \langle \bullet \rangle$ denotes averaging with respect to different disorder realization. 
	
	The  model of disorder that we consider  here is the speckle potential, which is widely used in experimental investigations of disordered BECs. 
	The speckle pattern is generated by deflecting a laser beam through a rough plate. 
	It can be modeled by \cite{Sanchez}:
	\begin{center}
		\begin{equation}
			U_{\text{dis}} ({\bf r})=U_{0}\displaystyle\sum_{i=1}^{S} U( {\bf r}- {\bf r}_{i}),  
		\end{equation} 
	\end{center}
	where $U_{0}$ is the strength of the disorder, and $r_{i}$ are the uncorrelated random positions, $S$ is the number of impurities, 
	and $U$ is a real valued-function of width $\sigma$ and has Gaussian shaped impurities $U(r)=e^{-r^2/\sigma^2}/(\sigma \pi^{1/2})^3$.
	Optical speckles  are long-ranged potentials. Another important feature is that their amplitude, geometry and correlation length can be tuned.

	For our purposes, we will use a dimensionless GPE.
	To this end, we express spatial variable $r$, time $t$, and energy in terms of transverse harmonic oscillator units $l= \sqrt{\hbar/m\omega_0}$, $\omega_0^{-1}$, 
	and $\hbar \omega_0$, respectively. This gives
	\begin{equation} \label{dimGPE}
		i\dfrac{\partial\phi}{\partial t}=\bigg(-\dfrac{1}{2}\nabla^2+\dfrac{1}{2} r^2-  \frac{4\pi a}{l} |\phi|^2+\tilde U_{0}\displaystyle\sum_{i=1}^{S} e^{-(r-r_{i})^2/\sigma^2}\bigg)\phi,
	\end{equation}
	where   $\tilde U_{0}=U_{0}/\hbar \omega_0(\sigma \pi^{1/2})^3$.
	
	It is also useful to write the functional energy  corresponding to GPE (\ref{dimGPE}) as:  
	\begin{align} \label{degy}
E[\phi^*, \phi]= \int\Bigg[\frac{1}{2}|\nabla\phi|^2+\dfrac{1}{2} r^2 |\phi|^2+\tilde U_{0}\displaystyle\sum_{i=1}^{S} e^{-(r-r_{i})^2/\sigma^2}|\phi|^2 -\frac{2\pi a}{l}|\phi|^4\Bigg] d {\bf r},
	\end{align}
	where the first term in the r.h.s represents the kinetic energy, the second term accounts for the potential energy, the third term is the disorder contribution to the energy, 
	and the fourth term can be regarded as the interaction energy.

	\section{Variational method}

	The GPE (\ref{dimGPE}) cannot be solved analytically, and even the numerical solution is difficult in the presence of negative interactions and disordered potentials. 
	It is therefore reasonable to attempt approximate solutions such as a variational scheme, in order to obtain a qualitative understanding of the  behavior of attractive BECs. 
	The variational method allows us to determine the ground-state energy and the wavefunction of the system by introducing a trial wavefunction that depends on one or more variational   parameters.

	\subsection{Equilibrium state}
	In the presence of a harmonic trap,  the sensible choice for a trial solution is a Gaussian function with variable width, $q$:
	\begin{equation} \label{Var}
		\phi({\bf r})=\sqrt{\frac{N}{\pi^{3/2} q^3}} \exp{\bigg(\frac{-r^2}{2q^2}}\bigg),
	\end{equation} 
	the validity of the ansatz (\ref{Var}) requires that the number of particles to be small, $N < N_{\text{cr}}$, where $N_{\text{cr}}$ is the critical atoms number.
	In such a limit the interaction term becomes negligible in the equation.
	
	Inserting the ansatz (\ref{Var}) into Eq.~(\ref{degy}), we get for the dimensionless functional energy  
	\begin{widetext}
		\begin{align} \label{degy1}
			\frac{E}{N}&=\frac{3}{4q^2}+\frac{3q^2}{4}-\frac{\eta}{\sqrt{2\pi}q ^3}+\displaystyle\sum_{i=1}^{S}\frac{\tilde U_{0}}{(\sigma^2+q^2)^{3/2}}\bigg[\frac{2r_{i}q\sigma^2}{\sqrt{\pi}}(\sigma^2+q^2)^{-1/2}\exp {\left(-\frac{r_{i}^2}{\sigma^2}\right)} \nonumber\\&+\bigg(\sigma^3+\frac{2r_{i}^2q^2\sigma}{(\sigma^2+q^2)}\bigg)\exp{\left(-\frac{r_{i}^2}{\sigma^2+q^2}\right)}\bigg(2-\text{erfc}\bigg(\frac{r_{i}}{\sigma}\bigg(\frac{\sigma^2}{q^2}+1\bigg)^{-1/2}\bigg)\bigg)\bigg], 
		\end{align}
	\end{widetext}
	where $\eta =N a/l$, and $\text{erfc}(r)$ is the complementary error function.
	
	\begin{figure}
	\begin{center}
	\centering
	\includegraphics[scale=0.8]{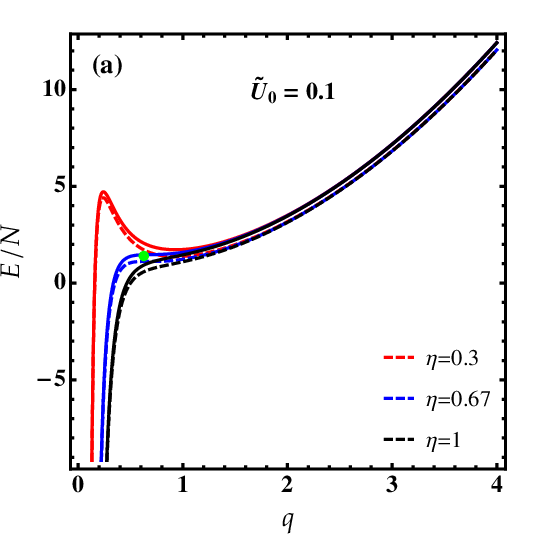}
	\includegraphics[scale=0.8]{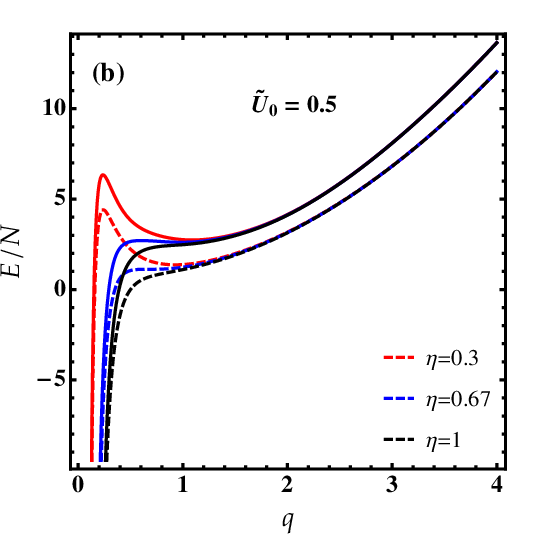}
	\caption{Energy  from equation (\ref{degy1}) as a function of the width $q$ for various values of the interaction strength $\eta$.  
Parameters are: $\sigma=0.2$, (a) $\tilde U_0=0.1$ and (b) $\tilde U_0=0.5$. Dashed lines: clean BEC. Solid lines: disordered BEC.}
	\label{Varegy}
	\end{center}
	\end{figure}
	
	Figure \ref{Varegy}.a shows that the energy has a maximum at very small $q$ and a minimum at $q=q_0 \simeq 1$ for small $\eta$.
	The two solutions start to disappear at a saddle (inflection) point marked with a dot in the figure.  In such a situation the 
        attraction force can overwhelm the repulsion, causing the condensate to collapse.
        Therefore, the critical value of the interaction parameter above which the disordered BEC collapses reads $\eta=\eta_{\text {cr}}\simeq 0.8$ which is larger by $\sim 13\%$  than that found
        in the absence of the disorder potential, $\eta_{\text {cr}}=0.671$. For ${}^7$Li atoms parameters: $a/l_0 \approx 5 \times 10^{-4}$ \cite{Hulet,Yong}, $\sigma=0.2$, and $\tilde U_0=0.1$,
one has for the critical atom number  $N_{\text {cr}}= \eta_{\text {cr}} l/a = 1600$.

	Remarkably, the disorder plays a dominant role only in the region where the size of the condensate is less than the harmonic oscillator length (i.e. $q < 1$)
	as it can be seen in Fig.~\ref{Varegy} b . It reduces the depth of the local minimum and increases the height of maximum.
	This clearly shows that the presence of the disorder may halt the collapse leading to stabilize the system. 
       This is in stark contrast with disordered ultracold bosons with repulsive interactions.

\begin{figure}
	\begin{center}
	\centering
	\includegraphics[scale=0.8]{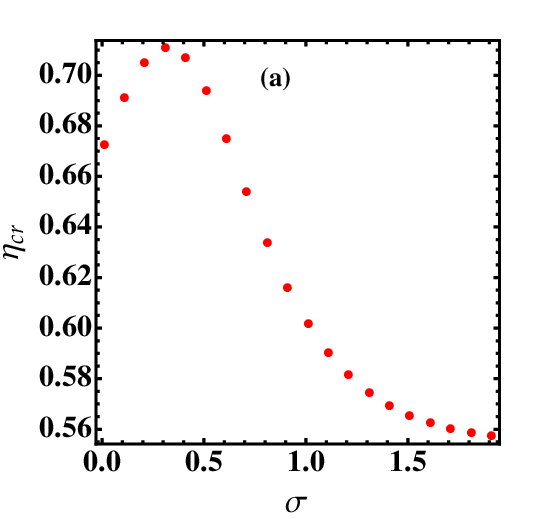}
	\includegraphics[scale=0.8]{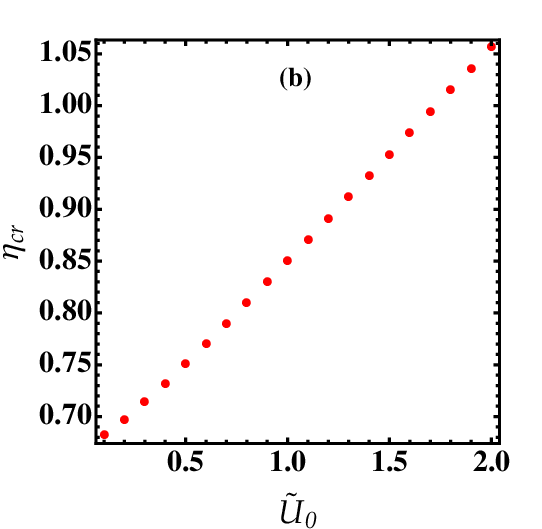}
	\caption{(a) Critical interaction strength $\eta_{\text {cr}}$ as a function of  $\sigma$ for $\tilde U_0=0.1$.
 (b) Critical interaction strength $\eta_{\text {cr}}$ as a function of $\tilde U_0$ for $\sigma=0.2$.}
	\label{CrSt}
	\end{center}
	\end{figure}	

To find the critical interaction strength $\eta_{\text {cr}}$,  one should minimize  the energy  (\ref{degy1}) with respect to the width, $q$, and then we solve the resulting equation numerically. 
In Fig.~\ref{CrSt} we plot $\eta_{\text {cr}}$ for which the local minimum of the energy disappears in terms of the disorder correlation length, $\sigma$, and the disorder strength, $\tilde U_0$.
We see that $\eta_{\text {cr}}$ increases sharply for $\sigma \lesssim 0.3$, it reaches its maximum at  $\sigma \approx 0.3$, then it decreases for $\sigma > 0.3$ where
the disorder effects become unimportant as is seen in Fig.~\ref{CrSt} a. Figure \ref{CrSt}~b shows that $\eta_{\text {cr}}$ increases linearly with the disorder strength $\tilde U_0$.
Another important remark is that for $\tilde U_0=0$ (i.e. in the absence of disordered effects), $\eta_{\text {cr}}$ well coincides with that derived for a clean BEC, $\eta_{\text {cr}} \simeq 0.67$.

	\subsection{Dynamics}

	This section deals with the dynamical properties of attractive disordered BECs. We use a variational method based on the following extended  Gaussian ansatz:
	\begin{equation} \label{Var1}
		\phi({\bf r}, t)=\sqrt{\frac{N}{\pi^{3/2} q^3(t) }} \exp\bigg({\frac{-r^2}{2q^2(t)}+i\theta(t)+i\gamma(t)r^2}\bigg),
	\end{equation}
	where the variational parameters are the condensate  width $q$, the chirp, $\theta$, and the phase, $\gamma$. The normalization factor ensures the conservation of the condition:
	\begin{equation}
		\int|\phi( {\bf r},t)|^2d^3 r=N.
	\end{equation} 
	The Lagrangian density reads:
	\begin{equation} \label{Larg}
		\mathcal{L}=\frac{i}{2}\bigg(\phi\frac{\partial\phi^*({\bf r}, t)}{\partial t}-\phi^*\frac{\partial\phi({\bf r}, t)}{\partial t}\bigg)+\frac{1}{2}|\nabla\phi({\bf r}, t)|^2+V({\bf r})|\phi({\bf r}, t)|^2
		+U_{\text{dis}}|\phi({\bf r}, t)|^2-\frac{2\pi a}{l}|\phi({\bf r}, t)|^4.
	\end{equation}
	Inserting the ansatz (\ref{Var1}) into Eq.~(\ref{Larg}) and integrating over space, we obtain for the Lagrangian of the system  $L=\int_{0}^{+\infty} d^3 { r} \mathcal{L}$:
	\begin{align} \label{Lag}
		L&=N \bigg(\dot{\theta}+\frac{3}{2}\dot\gamma q^2\bigg)+\frac{3}{4}N\bigg(4\gamma^2 q^2+\frac{1}{q^2}\bigg)+\frac{3}{4} N q^2
		-\frac{\eta N}{\sqrt{2\pi}q ^3}+\displaystyle\sum_{i=1}^{S}\frac{N \tilde U_{0}}{(\sigma^2+q^2)^{3/2}}\bigg[\frac{2r_{i}q\sigma^2}{\sqrt{\pi}}(\sigma^2+q^2)^{-1/2}\nonumber\\&\exp{\left(-\frac{r_{i}^2}{\sigma^2}\right)}+\bigg(\sigma^3+\frac{2r_{i}^2q^2\sigma}{(\sigma^2+q^2)}\bigg)\exp{\left(-\frac{r_{i}^2}{\sigma^2+q^2}\right)}\bigg(2- \text{erfc}
		\bigg(\frac{r_{i}}{\sigma}\bigg(\frac{\sigma^2}{q^2}+1\bigg)^{-1/2}\bigg)\bigg)\bigg].
	\end{align}
	The equations of motion can be derived from  the Euler-Lagrange equations  $d/d t\big(\partial L/\partial \dot Z \big)- \big(\partial L/\partial Z\big)$ 
	for variational parameters $Z \rightarrow q, \theta, \gamma$. Using the  Lagrangian (\ref{Lag}), we get
\begin{equation} \label{eqm0}
		\dot N=0,
	\end{equation}
	\begin{equation} \label{eqm1}
		3N\dot q q- 6N\gamma q^2=0,
	\end{equation}
	and 
	\begin{equation} \label{eqm2}
		3N\dot\gamma q+6N\gamma^2 q+\frac{d f}{d q}=0, 
	\end{equation}
	where    
	\begin{align}
		f( q)&=\frac{3}{4}\frac{ N}{q^2}+\frac{3}{4} N q^2-\frac{\eta N}{\sqrt{2\pi}q ^3}+\displaystyle\sum_{i=1}^{S}\frac{N \tilde U_{0} }{(\sigma^2+q^2)^{3/2}}\bigg[\frac{2r_{i}q\sigma^2}{\sqrt{\pi}}(\sigma^2+q^2)^{-1/2}\exp {\left(-\frac{r_{i}^2}{\sigma^2}\right)}\\& +\bigg(\sigma^3+\frac{2r_{i}^2q^2\sigma}{(\sigma^2+q^2)}\bigg)\exp{\left(-\frac{r_{i}^2}{\sigma^2+q^2}\right)}\bigg(2- \text{erfc}
		\bigg(\frac{r_{i}}{\sigma}\bigg(\frac{\sigma^2}{q^2}+1\bigg)^{-1/2}\bigg)\bigg)\bigg]. \nonumber 
	\end{align}
Equation (\ref{eqm0}) shows the conservation of the number of particles.
	Combining Eqs.~(\ref{eqm1}) and (\ref{eqm2}), we obtain
	\begin{equation}  \label{eqm11}
		\gamma=\dfrac{1}{2 q}\dfrac{d q}{d t},
	\end{equation}
	and 
	\begin{equation} \label{eqm22}
		\frac{d^2q}{d t^2}=-\frac{2}{3N}\frac{d f}{dq}\equiv -\frac{d U_{\text{eff}}}{d q},
	\end{equation}
	where 
	\begin{align}  \label {Upot}
		U_{\text{eff}} (q)&=\frac{1}{2q^2}+\frac{1}{2} q^2-\frac{\sqrt{2}}{3\sqrt{\pi}}\frac{\eta}{q^3}+\frac{2}{3}\displaystyle\sum_{i=1}^{S}\frac{ \tilde U_{0}}{(\sigma^2+q^2)^{3/2}}
		\bigg[\frac{2r_{i}q\sigma^2}{\sqrt{\pi}}(\sigma^2+q^2)^{-1/2} \exp{\left(-\frac{r_{i}^2} {\sigma^2} \right)} \\ 
		&+\bigg(\sigma^3+\frac{2r_{i}^2q^2\sigma}{(\sigma^2+q^2)}\bigg)\exp{\left(-\frac{r_{i}^2}{\sigma^2+q^2}\right)}\bigg(2- \text{erfc}
		\bigg(\frac{r_{i}}{\sigma}\bigg(\frac{\sigma^2}{q^2}+1\bigg)^{-1/2}\bigg)\bigg)\bigg], \nonumber
	\end{align}
	is the effective potential. 
	
\begin{figure}
	\begin{center}
	\centering
        \includegraphics[scale=0.8]{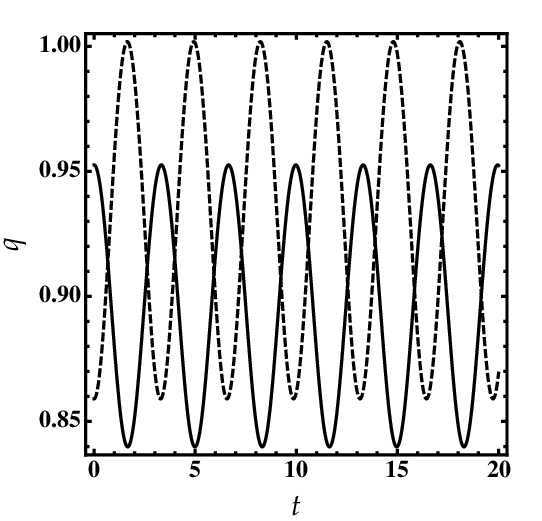}
	\caption{ Time evolution of the condensate width from Eq.~(\ref{eqm22}) for different values of the disorder strength $\tilde U_0$. 
 Solid line: $\tilde U_0=0.2$. Dashed line: $\tilde U_0=0$.
Parameters are  $\eta = 0.3$, and $\sigma=0.2$.}
	\label{TEW}
	\end{center}
	\end{figure}

	The time evolution of the condensate width is computed by a numerical solution of Eqs.~(\ref{eqm22}) and (\ref{Upot}) using the predicted equilibrium width as an initial condition.
	The results are shown in Fig.~\ref{TEW}.
	We observe that the width  exhibits periodic oscillations regardless of the value of the disorder strength. 
	The presence of the disorder potential leads to modify the amplitude and the width of such oscillations.

	\subsection{Breathing modes}
	
\begin{figure}
	\begin{center}
	\centering
	\includegraphics[scale=0.8]{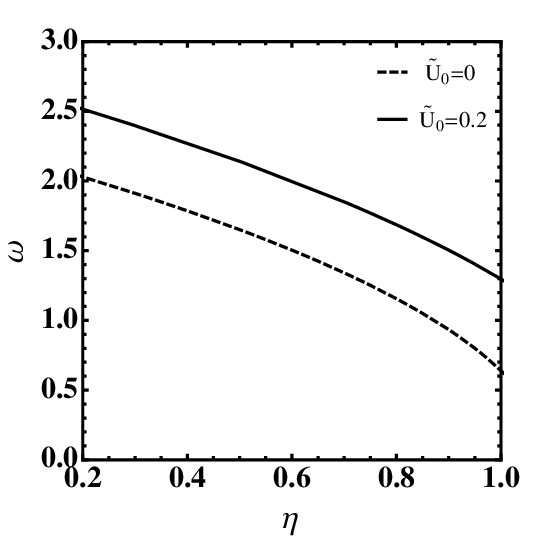}
	\caption{ Breathing mode oscillation frequency as a function of the interaction parameter $\eta$ for various values of the disorder strength $\tilde U_0$ and for $\sigma=0.2$.}
	\label{bmod}
	\end{center}
	\end{figure}

The low-lying excitations around the equilibrium solutions, $q_{0}$, can be evaluated employing the linearization $q(t) = q_{0} + \delta q(t)$ \cite{Abbas}, where $\delta q(t)\ll q_0$, and 
	$\delta q(t) = \delta q e^{i\omega t}$. This gives for the frequencies of the breathing modes:
	\begin{equation}
		\omega^2=\dfrac{d^2 U_{\text{eff}}}{d q^2}\vert_{q=q_{0}}.
	\end{equation}
	This yields
	\begin{equation} \label{feq}
		\omega=\sqrt{1-\frac{4\eta\sqrt{2/\pi}}{q_{0}^5}+\frac{3}{q_{0}^4}+\frac{2\tilde U_{0}}{3 \left(\sigma ^2+q _0^2\right){}^{7/2}} \sum_{i=1}^S\Big[U_{1}
			+U_{2}\Big]},
	\end{equation}
	where
	\begin{align*}
		U_{1}&= \left(e^{-\frac{r_i^2}{q_{0}^2+\sigma ^2}} \left(\frac{2q_{0}^2 \sigma  r_i^2}{q_{0}^2+\sigma ^2}+\sigma ^3\right)
		\left(2-\text{erfc}\left(\frac{r_i}{\sigma  \sqrt{\frac{\sigma ^2}{q_{0}^2}+1}}\right)\right)+\frac{2 q_{0} \sigma ^2 r_i e^{-\frac{r_i^2}{\sigma
					^2}}}{\sqrt{\pi } \sqrt{q_{0}^2+\sigma ^2}}\right)\left(12 q_{0}^2-3 \sigma ^2\right)\\ &-\frac{12q_{0}r_i
			e^{-\frac{r_i^2}{q_{0}^2+\sigma ^2}} \left(\frac{2q_{0}^2 \sigma  r_i^2}{q_{0}^2+\sigma ^2}+\sigma ^3\right) \left(q_{0} r_i
			\left(2-\text{erfc}\left(\frac{r_i}{\sigma  \sqrt{\frac{\sigma ^2}{q_{0}^2}+1}}\right)\right)+\frac{\sigma  \sqrt{q_{0}^2+\sigma ^2} e^{-\frac{r_i^2}{\sigma
						^2 \left(\frac{\sigma ^2}{q_{0}^2}+1\right)}}}{\sqrt{\pi }}\right)}{\left(q_{0}^2+\sigma ^2\right)}\\ &-\frac{12q_{0} \sigma ^3 r_i 
			\left(2q_{0}r_i\sqrt{\pi }\left(q_{0}^2+\sigma ^2\right)^{-1/2}	e^{-\frac{r_i^2}{q_{0}^2+\sigma ^2}} \left(2-\text{erfc}\left(\frac{r_i}{\sigma  \sqrt{\frac{\sigma ^2}{q_{0}^2}+1}}\right)\right)+ \sigma    e^{-\frac{r_i^2}{\sigma ^2}}\right)}{\sqrt{\pi } (q_{0}^2+\sigma ^2)^{1/2}},
	\end{align*}
	and
	\begin{align*}
		U_{2}&=\frac{16 r_i^3\sigma ^3 e^{-\frac{r_i^2}{q_{0}^2+\sigma ^2}}\left(q_{0}^2 r_i \left(2-\text{erfc}\left(\frac{r_i}{\sigma  \sqrt{\frac{\sigma
						^2}{q_{0}^2}+1}}\right)\right)+\sigma  \sqrt{q_{0}^2+\sigma ^2} e^{-\frac{r_i^2}{\sigma ^2 \left(\frac{\sigma ^2}{q_{0}^2}+1\right)}}\right)}{\sqrt{\pi }\left(q_{0}^2+\sigma ^2\right)^2}+\frac{4q_{0} \sigma  r_i^3 e^{-\frac{r_i^2}{q_{0}^2+\sigma ^2}-\frac{r_i^2}{\sigma ^2 \left(\frac{\sigma ^2}{q_{0}^2}+1\right)}} \left(\frac{2 q_{0}^2
				\sigma  r_i^2}{q_{0}^2+\sigma ^2}+\sigma ^3\right)}{\sqrt{\pi } \left(q_{0}^2+\sigma ^2\right)^{3/2}}\\&-\frac{2 \sigma  r_i^2 e^{-\frac{r_i^2}{q_0^2+\sigma ^2}} \left(2 r_i^2 \left(-2 q_0^2 \sigma ^4+q_0^4 \sigma ^2+3 q_0^6\right)-4 q_0^4
			r_i^4+\left(q_0^2+\sigma ^2\right){}^2 \left(13 q_0^2 \sigma ^2+2 q_0^4-\sigma ^4\right)\right) \left(2-\text{erfc}\left(\frac{r_i}{\sigma 
				\sqrt{\frac{\sigma ^2}{q_0^2}+1}}\right)\right)}{\left(q_0^2+\sigma ^2\right){}^3}\\&-\frac{6 q_{0} \sigma ^2 r_i  \left(e^{-\frac{r_i^2}{q_{0}^2+\sigma ^2}-\frac{r_i^2}{\sigma ^2 \left(\frac{\sigma ^2}{q_{0}^2}+1\right)}}
			\left(\frac{2 q_{0}^2 r_i^2}{q_{0}^2+\sigma ^2}+\sigma ^2\right)+\sigma ^2 e^{-\frac{r_i^2}{\sigma ^2}}\right)}{\sqrt{\pi } \left(q_{0}^2+\sigma ^2\right)^{1/2}}.
	\end{align*}
       The dependence of $\omega$ on $\eta$ is shown in Fig.~\ref{bmod}. 
	We clearly see that the frequency of the breathing modes decreases with increasing the interaction strength and thus with the number of particles.
        The same figure shows that $\omega$ increases with large disorder strength, $\tilde U_0$.
	Importantly, the frequency approaches zero as the condensate becomes unstable with respect to compression. 
	
	\section{Numerical results}
	
	In order to gain additional insights into the behavior of disordered attractive BEC, we perform the numerical integration of the static GPE 
        which can be derived by introducing the transformation: $\phi({\bf r},t)=\phi({\bf r}) \exp{(-i\mu t)}$ with $\mu$ being the chemical potential, into Eq.~(\ref{dimGPE}).
       Many numerical methods have been proposed to obtain solutions for GPE (see e.g \cite{Adh,Bao, Arnold,Altm} and references therein).
        Here, we adopt the split-step Fourier spectral method \cite{Turi,YS} with space step 0.04, and time step 0.01. The Gaussian ansatz  (\ref{Var}) is used as an initial wavefunction.  
       The validity and accuracy of the results are checked by varying the space and time steps and the total number of space and time steps \cite{YS}. The time evolution is  continued till convergence.
	To generate the speckle potential we use a set of random numbers which are then mapped into the interval $[0,L]$ by a linear transformation \cite{Sanchez,YS,Abbas}.
	We choose $S=300$, $L=30$, and a small width, $\sigma =0.2$.

\begin{figure}
	\begin{center}
	\centering
	\includegraphics[scale=0.85]{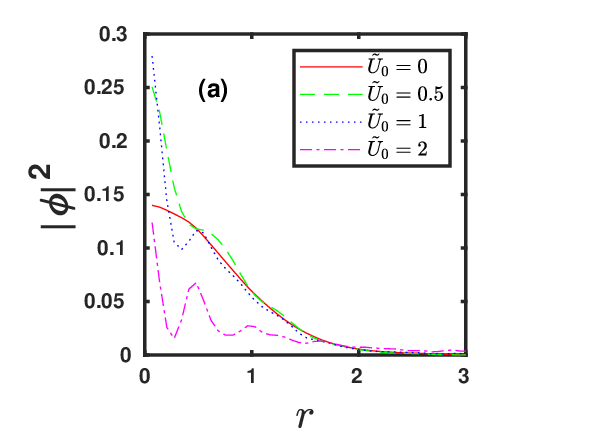}
        \includegraphics[scale=0.85]{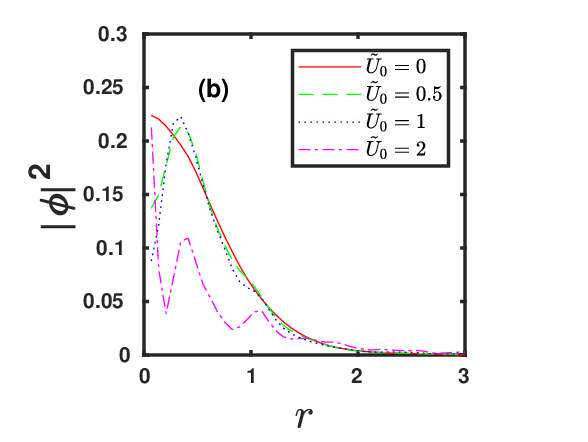}
	\caption{ (a) Density profiles of a disordered attractive BEC for different values of the disorder strength obtained by solving numerically the stationary GPE (\ref{dimGPE}).
Parameters are:  $\eta=0.3$, $q = 1$ and $\sigma=0.2$.
(b) The same as subfigure (a) but for  $\eta = 0.8$.}
	\label{D2}
	\end{center}
	\end{figure}

\begin{figure}
	\begin{center}
	\centering
	\includegraphics[scale=0.9]{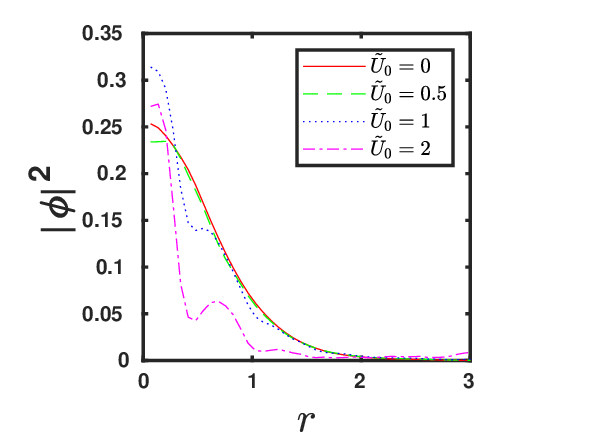}
	\caption{ The same as in Fig.~\ref{D2} but for  $\eta =1$.}
	\label{D3}
	\end{center}
	\end{figure}

Remarkably, in the absence of disorder, the shape of the condensate is smooth in good agreement with the results of clean BECs \cite{Hulet1,Dodd}. 
This clearly shows the viability and fiability of our numerical scheme.
For weak disorder $(\tilde U_0 < 0.5)$, the condensate remains extended and phase coherent (i.e. the resulting density does not fluctuate) as is seen in Fig.~\ref{D2} (a).
However,  moderate disorder may  induce density modulations in the condensate.
The appearance of such distortions in density can be attributed to the interplay of attractive interactions and disorder.

Our numerical results suggest also that before the condensate reaches the collapse region ($\eta<\eta_c$), these density modulations are increasing 
with the disorder strength and with the interactions strength $\eta$ as shown in Fig.~\ref{D2} (b). 
For sufficiently strong disorder ($\tilde U_0 \gtrsim 2$), however, the BEC may fragment into several peaks.
In such a situation, the disorder dominates the interaction energy which is converted into kinetic energy giving rise to deform the condensate and eventually break it into many blocks.

Surprisingly, near the collapse region, $\eta=1$, the density modulations significantly reduce even for large $\tilde U_0$ (see  Fig.~\ref{D3}).
The  condensate width decreases while its amplitude increases with the disorder indicating that the number of atoms is continuously decreasing.

	\section{Conclusion} 

	In this paper we systematically studied the equilibrium and dynamical properties of attractive BECs confined in a harmonic trap subjected to a 3D optical speckle potential 
	with a combined variational and numerical methods.
        We calculated the functional energy, equations of motion for different variational parameters, the width and the frequency of the breathing modes of the condensate 
	in terms of the disorder parameters. The critical interaction strength has been also established and examined in terms of the strength and the width of the disorder potential. 
        We addressed in addition the collapse dynamics in different regimes. 
	Our numerical predictions revealed that the presence of a disorder potential may cause density modulations which are increasing with the disorder strength.
       This latter plays a major role in the stability of attractive BECs.
	One can infer that the disorder potential can be regarded as a new mechanism that may adjust the collapse of the condensate.

	\section*{Acknowledgements}
       We thank Evgeny Sherman for valuable discussions and comments.

\end{document}